\documentclass[12pt]{article}
\usepackage[dvips]{graphicx}
\usepackage{amsmath,amssymb} 
\begin{document}

%----------------------------------------- 
\title{Constraints on the charged Higgs boson of the two Higgs doublet model type III} 

\author{H. Cardenas$^{1,2}$, J. Duarte$^1$ and J.-Alexis Rodriguez$^1$ \\
1. \it Departamento de F\'{\i}sica, Universidad Nacional de Colombia \\  \it Bogot\'a, Colombia \\ 
2. \it Escuela de F\'{\i}sica, Universidad Pedag\'ogica y Tecnol\'ogica de Colombia \\
\it Duitama, Colombia}

\maketitle

\begin{abstract}
The D0 experiment has reported a direct search for a charged Higgs boson produced by $q \bar q$ annihilation 
and decaying to $t \bar b$ final state, in the $180 \leq M_{H^+} \leq 300$ GeV mass range.  
The analysis has lead to upper limits on the production cross section in the framework of the two Higgs doublet model  types I, II and III.  We 
compare the predictions of two different scenarios in the framework of the two Higgs doublet type III to the cross section limits reported by D0 
collaboration, and we obtain constraints on the charged Higgs 
mass, for the case when the charged Higgs mass is bigger than the top quark mass.  Also, searches for the charged Higgs boson with a mass smaller
 than top quark mass 
are considered,  we discuss the possible limits on the charged Higgs boson mass  obtained from measurements  of 
the ratio $R_\sigma=\sigma_{t \bar t}^{l+jets}/\sigma_{t \bar t}^{dileptons}$ within  the  two Higgs doublet model type III.

\end{abstract}

%\pacs{12.60.Fr, 14.80Cp, 13.85.Rm}

%------------------------------------------------------------- 
 
%\section{Introduction} 

The Two Higgs Doublet Model (2HDM) is a  simple extension of the standard model (SM), where an additional Higgs doublet is introduced \cite{higgshunter}.
The two  complex scalar fields
correspond to eight degrees of freedom, where three of them are identified as
Goldstone bosons and are absorbed as longitudinal components to the
$W^\pm$ and $Z$ bosons giving mass to the weak bosons. The remaining degrees of freedom are interpreted as five physical states: two neutral
scalars $h^0$ and $H^0$, a pseudo-scalar $A^0$, and a pair of charged
Higgs bosons $H^\pm$.  Another consequence of this extension is a more generic pattern of Flavor Changing Neutral Currents (FCNC), where 
such processes are included at tree level. Because of low energy experiments these FCNC at tree level could be a problem, usually solved by 
imposing a discrete symmetry that allows to couple at most one Higgs doublet to each fermion. In the 2HDM called type I, one Higgs doublet gives 
masses to the up and down quarks, simultaneously. And in the model type II, one Higgs doublet gives masses to the up type quarks and the other one to the
 down type quarks, which is precisely the structure of the minimal supersimmetric extension of the SM (MSSM). However, the addition of these discrete
 symmetries is not mandatory and, in that case, both doublets contribute to generate the masses for up-type and down-type quarks. Such model is known as
 the 2HDM type III \cite{original}. 

While it may be hard to distinguish any one of
these neutral Higgs bosons of the 2HDM from that one of the SM, the
charged $H^\pm$ pair carry a distinctive hall-mark of this kind of new physics. Hence the
charged Higgs boson plays a very important role in the search of new physics beyond the SM. Usually the limits on the charged Higgs mass have been
 studied in the context of the 2HDM type II. Direct searches have been carried out by LEP experiments, searching for pair produced charged Higgs bosons
 via $s$-channel exchange of a {\it{Z}}-boson or a photon. The LEP collaboration analysis yield a combined lower limit on $M_{H^\pm}$ of 78.6 GeV \cite{LEP} 
assuming $H^+ \to \tau^+ \nu_\tau(c \bar s)$ in a wide range of the ratio of the vacuum expectation values of the two 
Higgs doublets, $\tan \beta=v_2/v_1$. Searches of the charged Higgs boson distinguish two cases $M_{H^\pm} > m_t$ and $M_{H^\pm} < m_t$. For the first case, at the
 Tevatron, the direct searches for charged Higgs boson are based on $p \bar p \to t \bar t$ with at least one top quark decaying into $t \to H^+ b$. The
 CDF collaboration also has reported a direct search for charged Higgs boson decaying into  $\tau^+ \nu_\tau$, $c \bar s$ $t \bar b$ or $W^+ A^0$, which uses 
measurements of the top pair production cross section in the  $leptons + \not E_T +jets + leptons$ channels, from data samples corresponding to an integrated 
luminosity of $193$ pb$^{-1}$ \cite{CDF1}. On the other hand, recently the D0 collaboration has presented a direct search for a charged Higgs boson produced 
by $q \bar q$ annihilation and decaying to $t \bar b$ final state, in the $180 \leq M_{H^+} \leq 300$ GeV mass range,  using 0.93 fb$^{-1}$ of data collected 
at center-of-mass energy $\sqrt{s}=1.96$ TeV \cite{D01}.  The analysis leads to upper limits on the production cross section in the 2HDM types I, II and III 
\cite{D01,peters}. We present two different scenarios for the two Higgs doublet model type III taking into account the previous 
bounds obtained in the literature on the parameter space of the model \cite{ARS,radiaz,rozo}. We calculate two different scenarios and we compare them with the data reported by
 D0 collaboration, and we obtain some bounds for a charged Higgs mass in the range 180-300 GeV.   Besides, searches in the $M_{H^\pm} < m_t$ region have been 
performed, using the production cross section of top quark pairs at the Tevatron \cite{CDF2, CDFII,D0oct, D0dic,D0R}. For instance, CDF II \cite{CDFII} has searched a charged Higgs boson decaying into $c \bar s$, this is
the first attempt to search directly $H^+ \to c \bar s$, they have not found any evidence of a charged Higgs boson but instead they put upper limits on $B(t \to H^+ b)$ between 0.1 and 0.3
assuming $B(H^+ \to c \bar s)=1$. Also the D0 collaboration has searched a charged Higgs boson in top quark decays, they \cite{D0oct} have used 0.9 fb$^-1$ of data assuming the subsequent decay 
$H^+ \to \tau^+ \nu_\tau$ and they exclude branching fractions $B(t \to H^+ b)$ between 0.19 and 0.24 for charged Higgs boson mass between 155 and 80 GeV. Other D0 search 
\cite{D0dic} for a charged Higgs boson used 1 fb$^-1$ of data at $\sqrt{s}=1.96$ TeV and they got upper limits on $B(t \to H^+ b)$ considering different scenarios depending
 on the values of $B(H^+ \to c \bar s)$ and $B(H^+ \to \tau^+ \nu_\tau)$. They exclude $B(t \to H^+ b)>0.22$ if $B(H^+ \to c \bar s)=1$ for $M_{H^\pm}$ between 80 and 155 GeV, and
exclude $B(t \to H^+ b)>0.15-0.19$ if $B(H^+ \to \tau^+ \nu_\tau)=1$. Finally, D0 collaboration \cite{D0R}  has also used the ratios of $t \bar t$ cross sections in different final states
to set upper limits on the branching fractions $B(t \to H^+ b \to \tau^+ \nu_\tau b)$ and $B(t \to H^+ b \to c \bar s b)$ as a function of the charged Higgs boson mass; they summarize
this information in the ratio $R_\sigma^{ll/lj}=\sigma_{t \bar t}^{ll}/ \sigma_{t \bar t}^{lj}$ which we are going to use in this work.  We discuss the possible limits on the charged Higgs 
boson mass obtained from measurements  of the ratio $R_\sigma^{lj/ll}$ within the framework of the  2HDM
type III for the $M_{H^\pm} < m_t$ region. 

Other experimental bounds on the charged Higgs boson mass come from virtual charged Higgs boson production in $b\to s\gamma$ transitions
 \cite{bsgamma}. Finally, the search for the charged Higgs boson will continue above the top quark mass at the LHC with  ATLAS and CMS. The main 
production mechanisms would be the processes $gg \to tb H^+$ and $gb \to t H^+$ which have been studied using simulations of the LHC detectors \cite{LHC}.

\begin{figure}[htp]
\begin{center}
      \includegraphics[scale=0.3]{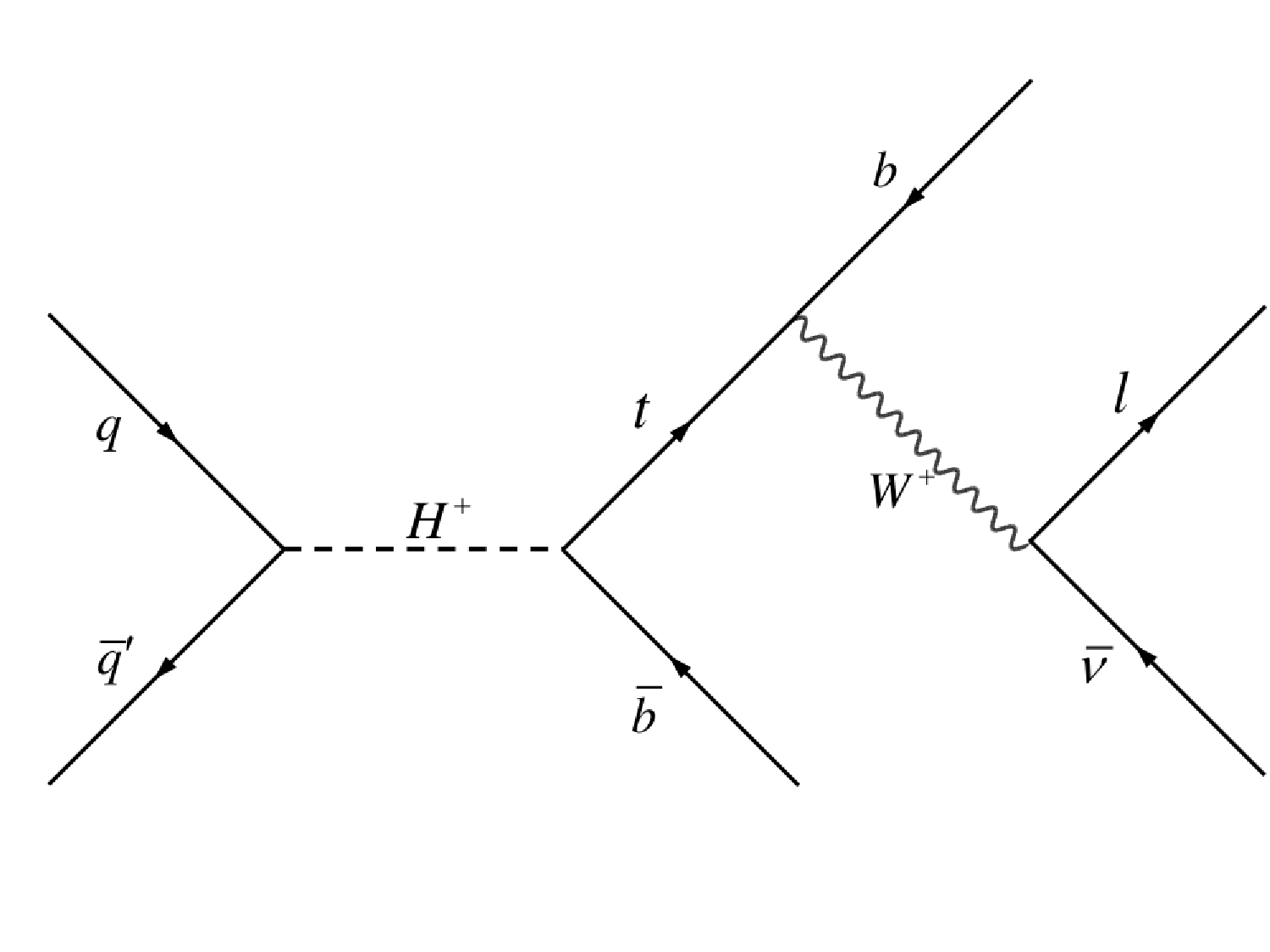} 
    \end{center}
\caption{The process $q \bar q' \to H^+ \to t \bar b$ followed by $t \to b (W^+ \to l \nu)$ }
\label{process}
\end{figure}

%\section{the two Higgs doublet model}
The 2HDM includes a second scalar doublet, with each doublet acquiring a vacuum expectation value (VEV), $v_i$, different from zero.
The scalar content of the model becomes:
\begin{equation}
\Phi_i=\left(\begin{array}{c}
 \phi_i^+\\
 \phi_i^0
\end{array}\right), \ \ \ \ 
\left<\Phi_i\right>=\left(\begin{array}{c}
 0\\
 \frac{v_i}{\sqrt2}
\end{array}\right), \ \ \ \
i=1,2 \,\, .
\end{equation}
In this way, the scalar spectrum of mass eigenvalues contains two CP-even neutral Higgs bosons $(h^0,H^0)$ coming from the mixing 
of the real part of the neutral components of both doublets with a mixing angle $\alpha$. Two charged Higgs bosons ($H^{\pm})$, which mix with 
the would-be Goldstone bosons ($G_W^{\pm}$) through the mixing angle $\tan\beta =v_2/v_1$ and one CP-odd Higgs ($A^0$), which mixes with the neutral
 would-be Goldstone.

\begin{table}[ph]

\begin{center}
%EndExpansion
\begin{tabular}{||c|c|c||}
\hline\hline
\textit{Bound} & \textit{Process} & \textit{Main assumption} \\ 
\hline
\multicolumn{1}{||c|}{$\xi _{\mu \tau }^{2}\in \left[ 7.62\times
10^{-4}\, , \,4.44\times 10^{-2}\right] $ \cite{radiaz}} & $\left(
g-2\right) _{\mu }$ & $m_{A^{0}}\rightarrow \infty $ \\ \hline
\multicolumn{1}{||c|}{$\xi _{\tau \tau }\in \left[ -1.8\times
10^{-2}\, , \,2.2\times 10^{-2}\right] $ \cite{radiaz}} & $\tau
^{-}\rightarrow \mu ^{-}\gamma $ & $m_{A^{0}}\rightarrow \infty $ \\ \hline
\multicolumn{1}{||c|}{$\xi _{\mu \mu }\in \left[ -0.12\, , \,0.12\right] $ \cite{radiaz}} & $\tau ^{-}\rightarrow \mu ^{+}\mu ^{-}\mu ^{-}$ & $%
m_{A^{0}}\rightarrow \infty $ \\ \hline
\multicolumn{1}{||c|}{$\xi _{\mu e}\in \left[ -0.39\, , \,0.39\right] $ \cite{radiaz}} & $\tau ^{-}\rightarrow e^{-}e^{-}\mu ^{+}$ & $%
m_{A^{0}}\rightarrow \infty $ \\ \hline
\multicolumn{1}{||c|}{$\lambda _{bb}\in \left[ -6\, ,\,6\right] $ \cite{jimenez}}
& $p\bar{p}\rightarrow b\bar{b}h\left( b\bar{b}\right) $ & $\cos \alpha \sim \sqrt{2}/2$ \\ \hline
\multicolumn{1}{||c|}{$\lambda _{tt}\in \left[ 0.5\, ,\, 1.7\right] $ \cite{bsgamma,stal}} & $B\rightarrow X_{s}\gamma $ & $M_{H^+}=250-500$ GeV \\ \hline
\multicolumn{1}{||c|}{$\left\vert \lambda _{tc}\right\vert \lesssim 2.3/\cos\alpha $ \cite{Milanes,larios}} & Electroweak precision observables & $m_{h}\lesssim 170 $ GeV \\ \hline
\multicolumn{1}{||c|}{$\xi_{ds,uc,db} \lesssim 10^{-5}$ \cite{ARS,wells}, $\xi_{bs} \lesssim 10^{-4}$ \cite{wells,joshipura}} & $F^0-F^0$ systems 
& $m_{h}=120$ GeV \\ \hline \hline
\end{tabular}%
\end{center}
\caption{\label{t:bounds}Experimental constraints over the $\xi_{ij}$ and $\lambda_{ij}$ parameters obtained by 
Refs. \cite{ARS},\cite{radiaz},\cite{bsgamma},\cite{Milanes},\cite{jimenez},\cite{larios},\cite{wells},\cite{joshipura},\cite{stal}. 
We should mention that restrictive values on $\lambda_{tt}$  and $\lambda_{bb}$ 
 are presented in a correlated form in reference \cite{stal}.}
\end{table}%

The most general Yukawa Lagrangian in this kind of models is
\begin{eqnarray}
-\mathcal L_Y&=&
\eta_{ij}^{U,0}\bar Q_{iL}^0\tilde\Phi_1 U_{jR}^0+\nonumber
\eta_{ij}^{D,0}\bar Q_{iL}^0 \Phi_1 D_{jR}^0\\
&+&\xi_{ij}^{U,0}\bar Q_{iL}^0\tilde\Phi_2 U_{jR}^0+
\xi_{ij}^{D,0}\bar Q_{iL}^0 \Phi_2 D_{jR}^0\nonumber\\
&+& \mbox{lepton sector\ +\ h.c}
\end{eqnarray}
where $\eta^{U,D}$ and $\xi^{U,D}$ are non-diagonal mixing matrices $3\times3$, $\tilde\Phi_i=i\sigma_2\Phi_i$, $(U,D)_R$ are 
right-handed fermion singlets, $Q_L$ are left-handed fermion doublets, and the index 0 indicates that the fields are not mass eigenstates.

In the general case, both Higgs doublets couple with the up and down sectors, and therefore they contribute 
simultaneously in the process of mass generation for quarks. This case leads to FCNC at tree level, because it is impossible 
to diagonalize simultaneously both matrices $\eta$ and $\xi$. This general case is known as 2HDM type III. However, FCNC processes 
at tree level are highly suppressed by the experiment. In order to avoid their existence,  the following set of discrete symmetries can be
 imposed:

\begin{eqnarray}
&\Phi_1\to\Phi_1 \mbox{ and } \Phi_2\to-\Phi_2,\nonumber\\
&D_{jR}\to\mp D_{jR} \mbox{ and } U_{jR}\to - U_{jR}.
\end{eqnarray}

The condition of invariance under one of these discrete symmetries leads to two cases:

\begin{itemize}
\item By using $D_{jR}\to -D_{jR}$, matrices $\eta^{U,0}$ and $\eta^{D,0}$ have to be eliminated from the Lagrangian.
 In this case $\Phi_1$ decouples in the Yukawa sector and only $\Phi_2$ gives masses to sectors up and down. This case is known as 2HDM type I.
\item By using $D_{jR}\to D_{jR}$, matrices $\eta^{U,0}$ and $\xi^{D,0}$ must be eliminated from the Lagrangian. In this case $\Phi_1$ couples to 
the down sector and $\Phi_2$ gives masses to up sector. This case is known as 2HDM type II.
\end{itemize}

\begin{figure}[hbp]
\begin{center}
	      \includegraphics[scale=0.5]{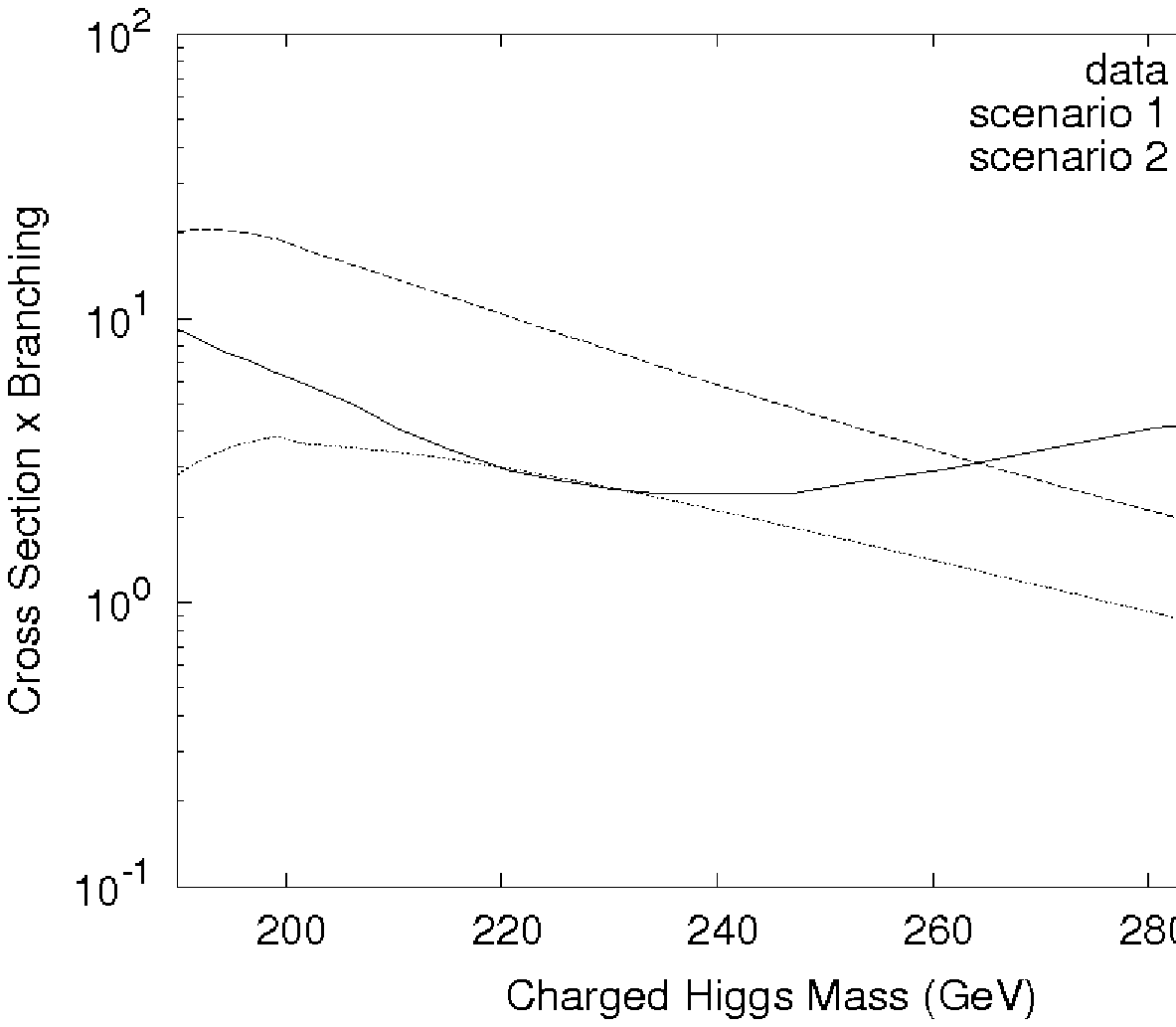} 
	    \end{center}
\caption{The cross section times branching fraction versus the charged Higgs boson for two different scenarios in the type III 2HDM }
\label{crossfig}
\end{figure}

In the framework of the type III 2HDM there is a global symmetry which can make a rotation 
of the Higgs doublets and fix one VEV equal to zero 
\cite{ARS,haber}. In such a way, $v_1=v$ and $v_2=0$, and the mixing 
parameter $\tan\beta=v_2/v_1$ can be eliminated from 
the Lagrangian.  If the parameter $\tan \beta$ is eliminated from the Lagrangian, we have the usual 2HDM type III \cite{ARS}, and  the Lagrangian of the charged sector is given by 
\begin{equation}
-{\cal L}^{III}_{H^\pm ud}=H^+ \bar U [ K \xi^D P_R-\xi^U K P_L] D+h.c.
\end{equation}
where $K$ is the Cabbibo-Kobayashi-Maskawa (CKM) matrix and $\xi^{U,D}$ the flavor changing matrices. 

For a better study of the FCNC processes, Cheng, Sher and Yuang (CSY) \cite{csy} proposed an anzats for the Yukawa matrices. It is
 based on the SM $\phi f\bar f$ couplings and states that 
$\xi_{ij}\equiv\frac{\sqrt{m_im_j}}{v}\lambda_{ij} \, .$ 
This is an ansatz for the Yukawa texture matrices looking for a phenomenological similarity
with SM couplings. This ansatz obeys to the fact that couplings between fermions and the Higgs particle in the SM are
 proportional to the mass of the fermion. Parameters $\lambda_{ij}$ could change the hierarchy between fermionic couplings and 
because of this it is expected that they would be $\sim 1$.  

Constraints on the parameters $\xi^{U,D,L}_{ij}$ can be found in the 
literature \cite{ARS,radiaz,rozo,bsgamma,Milanes,we,jimenez,larios,wells,joshipura,stal}. Constraints on $\xi^{U,D}_{ij}$ coming from $F^0-F^0$ mixing 
($F=K,B_d, B_s,D$) in a two Higgs doublet model type III where evaluated by Atwood, et.al \cite{ARS} and recently in reference \cite{wells} the bounds 
on $\xi_{ij}$ that contribute to these systems where updated using the latest experimental results on $\Delta M_F$. A discussion of $B^0_s-B_s^0$ system 
bounds and the processes $\bar B_s \to \mu^+ \mu^-$, $\bar B_d \to \bar K \mu^+ \mu^-$ was done in reference \cite{joshipura}. On the other hand, using $B$, 
 $K$ and $D_s$ meson decays constraints on $\lambda_{ii}$ are obtained in reference \cite{stal}; they also use $b \to s \gamma$ and $\Delta M_B$ to constraint 
the parameter $\vert \lambda_{tt} \vert \leq 1$ for $M_{H^+}=500$ GeV wich is consistent with the analysis of reference \cite{bsgamma}. And, studying the 
light Higgs boson production associated with $b$ quark production at Tevatron, ranges for the $\lambda_{bb}$ coupling were obtained in reference \cite{jimenez}.
Constraints on the parameter $\lambda_{tc}$ can be obtained from electroweak precision observables from LEP \cite{larios,Milanes}. Finally, bounds on the 
parameters in the leptonic sector $\xi_{ll'}$ have been obtained using the anomalous factor $(g-2)_\mu$ and tau decays in references \cite{radiaz,we}. A summary of the 
constraints on the parameters $\xi_{ij}$ is in table \ref{t:bounds}.

Recently, there is a renewed interest for the general two Higgs doublet model and there are a general class of solutions to the problem of 
the requirement that  the off-diagonal $\xi_{ij}$ should be extremely small in order to satisfy the FCNC constraints \cite{wells}. Many
models have been presented  in the literature specifying the Higgs boson interactions with the fermions in such way that tree-level
 FCNCs do not arise \cite{nuevos}.

%\section{The search of the charged Higgs boson at Tevatron}

The search for the charged Higgs boson by D0 collaboration \cite{D01} use the 
process $q \bar q' \to H^+ \to t \bar b$ followed by $t \to b (W^+ \to l \nu)$, which corresponds to single top production, with a
 final signature  $2b$-tagged $+ lepton +$ missing energy, see Fig. \ref{process}. We notice there are two vertices type $H^+ q \bar q'$ where 
the parameters $\xi_{q q'}$  appear and then different considerations can be done, taking into account that terms like $\sum_j (K_{qj} \xi_{j q'})$
 are present. In the annihilation vertex $q \bar q' H^+$ are possible the vertices $c \bar b H^+$ and $c \bar s H^+$ in the framework of the 2HDM type III.
 About this point H. He and C. P. Yuan \cite{yuan} showed that it is possible to enhanced the production cross section if the vertex $c \bar b H^+$ is 
relevant, and it will happen if the parameter $\lambda_{tc}$ is bigger than one. We have evaluated the option of the vertex $c \bar s H^+$ in the framework
 of the 2HDM-III and in fact, it is smaller two or three orders of magnitude than the vertex $c \bar b H^+$. Now in the second vertex, the $H^+$ decay vertex,
 we have several couplings regarding the 2HDM-III allowed processes: $H^+ \to t \bar s$, $H^+ \to  c \bar b$ and $H^+ \to t \bar b$. The first option, 
into $t \bar s$, does not have a quark $b$ in the final state,  and the experiment is looking for either one or two $b$-tags, one coming from the $H^+$ decay
 and the second one coming from the top quark decay.  The second option is  to decay into $c \bar b$ but it is doubly suppressed for the $\xi_{tc}$
 coupling and for the Cabbibo-Kobayashi-Maskawa $K_{cb}$ factor. Finally, we will have the same main channel production  in the 2HDM type III that 
in the types I and II, $c \bar c \to H^+ \to t \bar b$ followed by $t \to (W^+ \to l \nu) b$.

Using the parameterization CYS \cite{csy} and the values of the CKM matrix $K_{ij}$ elements, the flavor changing 
couplings of the charged Higgs boson to the fermions, which are proportional to $\sum_j K_{qj} \xi_{jq'}$ are reduced as is shown in Tab. \ref{acoples2}. These factors
 have been simplified taken into account the numerical values of quark masses and assuming that the parameters $\lambda_{ij}$ involving the 
first and second generations are smaller than parameters involving the third generation \cite{ARS}. Finally, bounds and restrictions on 
the $\lambda_{ij}$ for the quark sector and $\xi_{ij}$ for the leptonic sector can be found in table \ref{t:bounds}.

\begin{figure}[hbp]
\begin{center}
	     \includegraphics[scale=0.5]{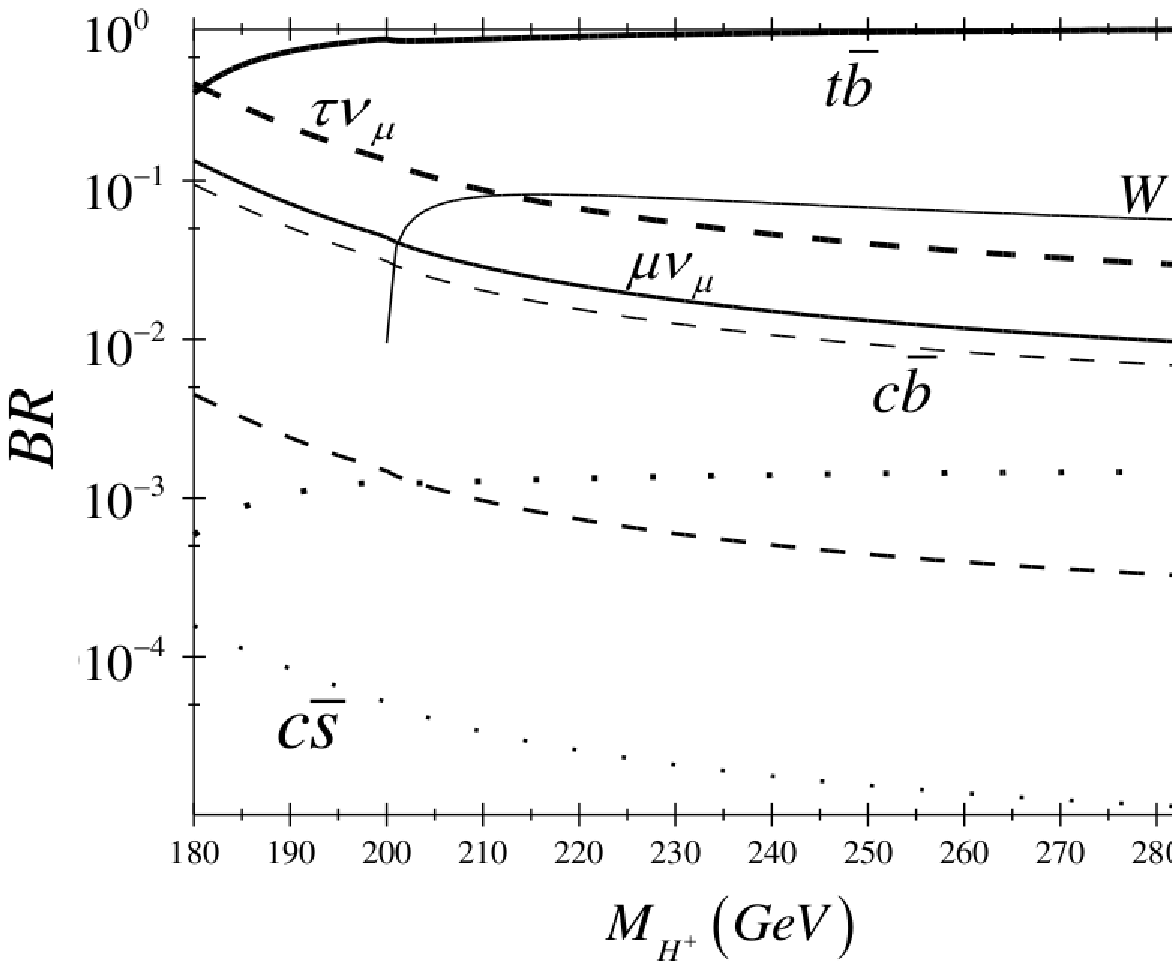} \quad
	 \includegraphics[scale=0.5]{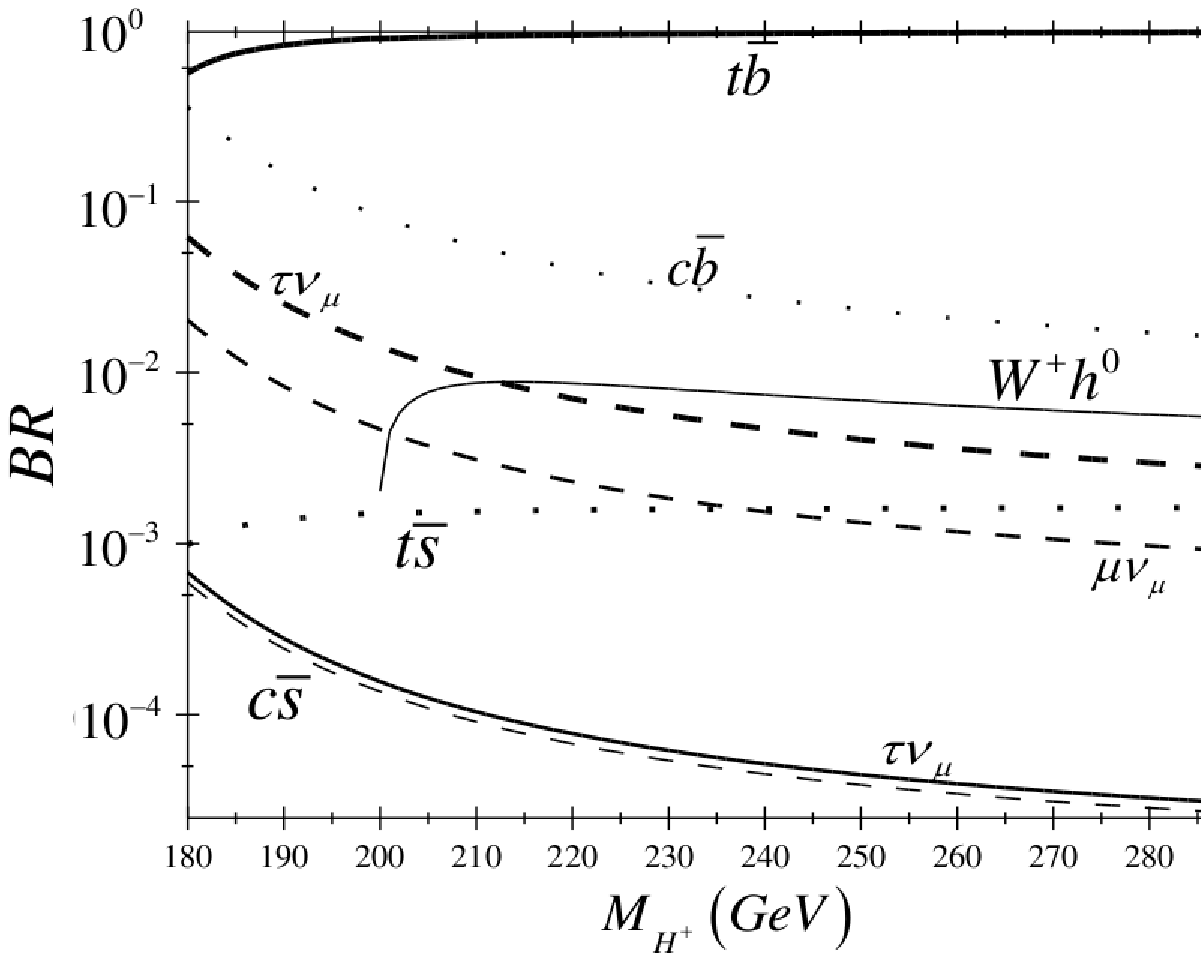} 
	    \end{center}
\caption{Branching ratios in
 the 2HDM-III for the parameters $\xi_{\tau \tau}=2.2 \times 10^{-2}$, $\xi_{\mu \tau}=2.1 \times 10^{-1}$, and $\xi_{\mu \mu}=0.12$ according
 to the allowed regions showed in table \ref{t:bounds}. 
The two scenarios are  $scenario \, 1: \lambda_{tc}=\lambda_{tt} =5, \lambda_{bb}=0$,  $scenario \, 2: \lambda_{tc}=3.5, \, 
\lambda_{tt} =0.5, \, \lambda_{bb}=4$. }
\label{branchings}
\end{figure}

\begin{table}[th]
\begin{center}
\begin{tabular}{||c|c|c||}\hline\hline
$(ij)$ & $(\xi^U K)_{ij}$ & $(K \xi^D)_{ij}$ \\ \hline
$c\overline{s}$ & $K_{ts}\frac{\lambda _{ct}\sqrt{m_{c}m_{t}}}{v}$ &  \\ \hline
$c\overline{b}$ & $K_{tb}\frac{\lambda _{ct}\sqrt{m_{c}m_{t}}}{v}$ & $K_{cb}
\frac{\lambda _{bb}m_{b}}{v}$ \\ \hline
$t\overline{s}$ & $K_{ts}\frac{\lambda _{tt}m_{t}}{v}$ &  \\ \hline
$t\overline{b}$ & $K_{tb}\frac{\lambda _{tt}m_{t}}{v}$ & $K_{tb}\frac{
\lambda _{bb}m_{b}}{v}$ \\ \hline\hline
\end{tabular}
\end{center}
\caption{\label{acoples2}Parameters involved in the flavor changing couplings of the charged Higgs boson.}
\end{table}

The analysis presented by D0 collaboration \cite{D01} for the type III 2HDM has followed the reference \cite{yuan} assuming that 
the parton level is important to enhance the cross section with $\lambda_{tc}$  bigger than one. 
The experimental analysis have used $\lambda_{tc}=5$ \cite{peters}. Further, they assumed the parameter $\lambda_{tt}$ in 
the charged Higgs decay vertex is equal to $\lambda_{tc}$. About this last point, we should mention that Atwood, {\it et. al.} in reference 
\cite{ARS} have already shown that the assumption $\lambda_{ij}=\lambda$ is not in agreement with the low energy phenomenology and on the other hand, 
it has been shown \cite{rozo} that perturbation theory validity requires that $\lambda_{tt} \leq 2.8$ but in any case,  Xiao and Guo \cite{bsgamma} have 
constrained the parameter $\vert \lambda_{tt} \vert < 1.7$ and for charged Higgs mass of the order of 200 GeV is favored
 $\lambda_{tt}\sim 0.5$. Finally in reference \cite{stal}, they have gotten limited regions in the plane $\lambda_{tt}-\lambda_{bb}$ using 
data from $\Delta M_{B_d}$ and other observables.
 From 
this point we aim to explore scenarios allowed 
in the 2HDM-III space parameters, with the additional simplification that in this specific model $\tan\beta$ is an spurious parameter \cite{haber}. The
 experimental
 D0 collaboration report the observed limits on the production cross section ($pb$) times the branching fraction
 $\sigma(q \bar q' \to H^+) \times B(H^+ \to t \bar b)$. These limits are shown in Fig. \ref{crossfig} labeled as data. We have
 used the program CompHEP \cite{comphep} to calculate the charged Higgs boson 
production and decays, $q \bar q' \to H^+ \to t \bar b \to W^+ b \bar b \to l^+ \nu b \bar b$ where $l$ represents an electron or muon. The expected 
limits using different values of the $\lambda_{ij}$ in the charged Higgs boson mass range 180 to 300 GeV are plotted in Fig. \ref{crossfig}. In
 Fig. \ref{crossfig}  the expected limits on $\sigma \times B$ are plotted using two scenarios: $\lambda_{tc}=5$, $\lambda_{tt}=5$ and  $\lambda_{bb}=0$
 which it is called here scenario 1  and 
the scenario 2 corresponds to  $\lambda_{tc}=3.5$, $\lambda_{tt}=0.5$ and  $\lambda_{bb}=4$. These  $\lambda_{ij}$ values in the scenario 2 are in
 the phenomenological
 allowed regions (see table 
\ref{t:bounds}) and basically
they are the upper value in the allowed interval in order to increase the possible effects coming from this specific model, the 2HDM-III. We conclude that restrictions on the parameter space
 of the type III 2HDM are not too strong. We notice that in the scenario 1, used by the D0 collaboration, only charged Higgs masses above
 around 264 GeV are allowed and for the scenario 2 the charged Higgs boson mass should be bigger than around  210-230 GeV. If we consider 
parameters $\lambda_{bb}$ and $\lambda_{tt}$ of the order $10^{-1}$ there is not gotten any constraint on the charged Higgs boson mass 
using the D0 data.

For completeness, the branching fraction ratios in the two scenarios considered are shown in Fig. \ref{branchings}, where we
 have used the parameters $\xi_{\tau \tau}=2.2 \times 10^{-2}$, $\xi_{\mu \tau}=2.1 \times 10^{-1}$, and $\xi_{\mu \mu}=0.12$ according to
 the allowed regions showed in table \ref{t:bounds}. The first scenario, 
reproduce the conditions for the experimental analysis done by Ref. \cite{D01}, 
where the dominant decay is $H^+ \to t \bar b$ followed by $H^+ \to c \bar b$ decay. In the second scenario, we have the final 
state $t \bar b$ as the dominant decay followed by the flavor changing decay into $\tau \bar \nu_\mu$ in the low mass region $M_{H^+}<210$ and for 
bigger masses  the decay into $W^+ h^0$ is the second probable decay.

\begin{figure}[htbp]
\begin{center}
	      \includegraphics[scale=0.5]{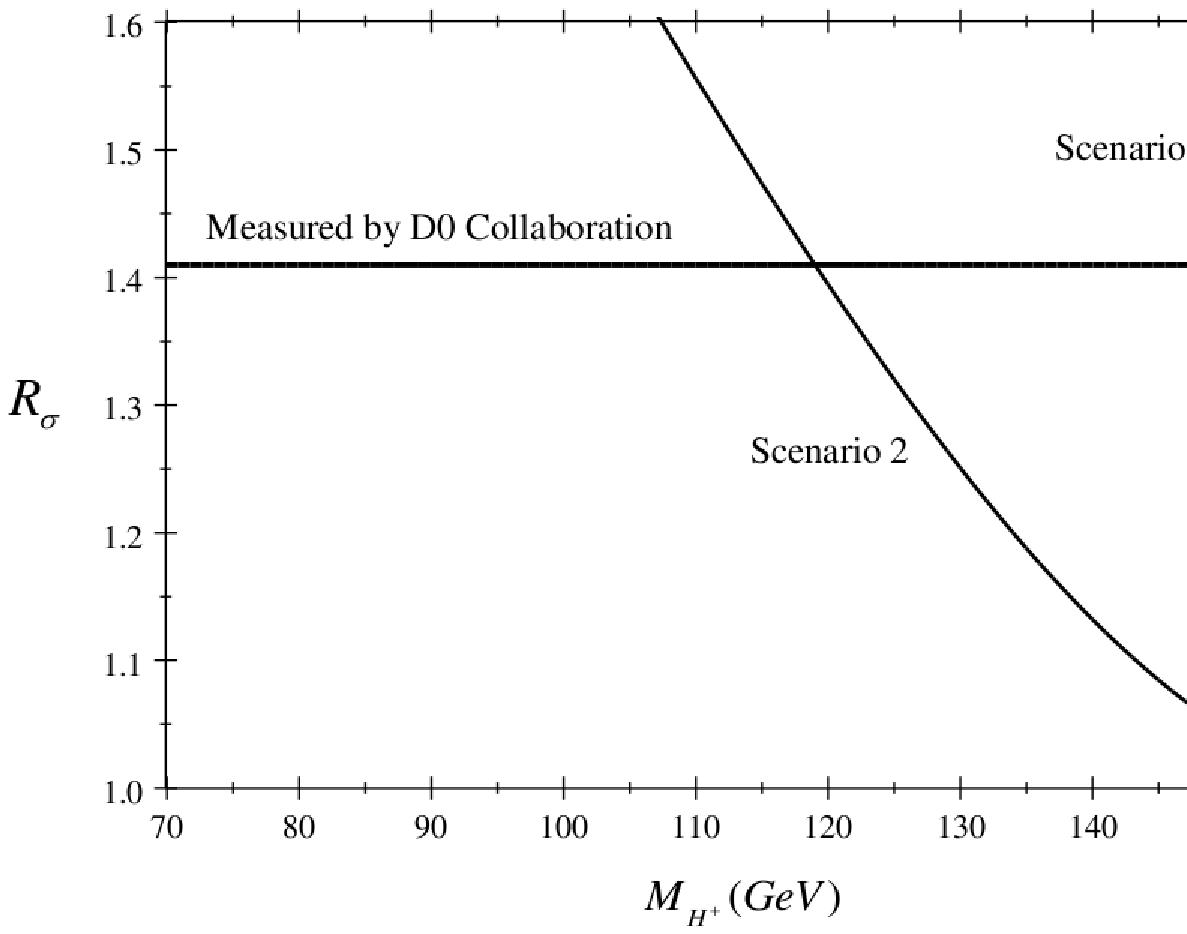} 
	    \end{center}
\caption{The ratio $R_\sigma^{lj/ll}$ with the experimental limits from D0 collaboration and the two scenarios specified in the text
 in the framework of the type III 2HDM }
\label{rsigma}
\end{figure}

The discussion presented until now is for charged Higgs boson mass bigger than the top quark mass,
 but we can also ask for the  case $M_{H^+}<m_t$. In this case the decay $t \to H^+ b$ competes with the standard decay 
channel $t \to W^+ b$. The experimental analysis has been done using the cross section of the top quark pair production \cite{peters,CDF2}. In order 
to estimate upper limits on the branching fraction $B(t \to H^+ b)$ it is useful
 to use the ratio $R_\sigma^{lj/ll}=\sigma(p\bar p \to t \bar t)_{l+jets}/\sigma(p \bar p \to t \bar t)_{dilepton}$. The ratio $R_{\sigma}$  should be
 consistent with one if the dilepton and lepton+jets analyses are in the framework of the standard model $t \bar t$ production. The
 ratio $R_\sigma$ will have smaller systematic uncertainties than individual cross section measurements and also some common factors will cancel, so 
it is a better place to look for new physics  rather than by comparing a measured cross section to a theory prediction. $R_\sigma$ is sensitive to
 decays such as $t \to H^+ b$. It is possible to give an interpretation of $R_\sigma$ in terms of the branching fraction $B(t \to H^+ b)$ and a 
measurement of this ratio can be interpreted as an upper limit on $B(t \to H^+ b)$.  In general this ratio can be written 
according to \cite{peters,quadt},
\begin{equation}
R_\sigma^{lj/ll} = 1 + \frac{B(t \to H^+ b)}{[1-B(t \to H^+ b)] B(W^+ \to q \bar q')} \, .
\end{equation}
For instance if we consider that $B(H^+ \to \tau^+ \nu_\tau)=1$ then $B(t \to H^+ b)=0.19-0.35$ for a charged Higgs boson of 80-155 GeV (see Fig 1 of \cite{D0R}).
Experimentally this ratio has been measured by D0 collaboration $R_\sigma^{lj/ll}=1.21^{+0.27}_{-0.26}$ \cite{peters} and more
recently reported $R_\sigma^{ll/lj}=0.86^{+0.19}_{-0.17}$ \cite{D0R} and by 
the CDF collaboration 
$R_\sigma^{ll/lj} =1.45^{+0.83}_{-0.55}$ \cite{CDF2}. We have plotted the ratio $R_\sigma^{lj/ll}$ in figure \ref{rsigma} in the framework of the 2HDM type III for the
 scenarios that we used so far. We obtain lower limits on the charged Higgs mass of $M_{H^+} \sim 150 $ GeV when
  $\lambda_{tc}=\lambda_{tt} =5, \, \lambda_{bb}=0$,  and $M_{H^+} \sim 120 $ GeV when  $\lambda_{tc}=3.5, \, \lambda_{tt} =0.5, \lambda_{bb}=4$.

As a conclusion we have reviewed the analysis presented by D0 collaboration \cite{D01} in the first search for a charged Higgs 
boson directly produced by quark-antiquark annihilation and decaying into the $t \bar b$ final state, in the $180 \leq M_{H^+} \leq 300$ GeV
 region in the framework of the 2HDM type III. The parameters of the two Higgs doublet model type III used by D0 collaboration on their report are 
out of the allowed ranges reported in literature, see table \ref{t:bounds}. 
  Two different scenarios for the two Higgs doublet model type III are presented.   We have calculated charged Higgs boson production and decays,
 $q \bar q' \to H^+ \to t \bar b \to W^+ b \bar b \to l^+ \nu b \bar b$ where $l$ represents an electron or muon.
 The expected limits on the $\sigma(q \bar q' \to H^+) \times B(H^+ \to t \bar b)$ for the two scenarios proposed are obtained and these limits allow 
a charged Higgs mass above around 264 GeV in the scenario 1: $\lambda_{tc}=\lambda_{tt} =5, \, \lambda_{bb}=0$ and bigger than 210-230 GeV for the
scenario 2: $\lambda_{tc}=3.5 \, \lambda_{tt} =0.5, \lambda_{bb}=4$.
  In addition, the branching fractions for the same  scenarios considered are shown in Fig. \ref{branchings}.  Finally, in the region $M_{H^+} < m_t$, we have used the
 reported  measurements of the ratio $R_\sigma$ \cite{peters,CDF2}. The lower limit
on the charged Higgs mass is $M_{H^+} \sim 150$ for the scenario 1 and  $M_{H^+} \sim 115$ for the 
scenario 2.

\section*{Acknowledgments}

 We would like to acknowledge to Y. Peters for useful discussions about the experimental results from her talk presented
 in SUSY08 conference Seoul, Korea, June 2008) and to C. Jimenez for his valuable help with data manipulation. Also we want to acknowledge 
to D. Milanes  and J. Idarraga for his careful reading of the manuscript and useful suggestions.

\end{document}